\documentclass[twocolumn,prb]{revtex4}
\usepackage{amsfonts}
\usepackage[T1]{fontenc}
\usepackage{amsmath,amsbsy,amssymb,graphicx}
\usepackage{times}
\usepackage{amsmath}
\usepackage{amssymb}
\usepackage{graphicx}
\usepackage{color}

\let\mathbf=\boldsymbol
\def\section#1{\medskip\noindent\textbf{#1}\par}

\begin{document}

\title{{\Large Magnetic bilayer-skyrmions without skyrmion Hall effect}}
\author{Xichao Zhang$^{1}$}
\author{Yan Zhou$^{1,2}$}
\email[Corresponding author:~]{yanzhou@hku.hk}
\author{Motohiko Ezawa$^{3}$}
\email[Corresponding author:~]{ezawa@ap.t.u-tokyo.ac.jp}
\affiliation{$^{1}$Department of Physics, University of Hong Kong, Hong Kong, China}
\affiliation{$^{2}$Center of Theoretical and Computational Physics, University of Hong Kong, Hong Kong, China}
\affiliation{$^{3}$Department of Applied Physics, University of Tokyo, Hongo 7-3-1, 113-8656, Japan}

\begin{abstract}\bf\noindent
Arising from emergent electromagnetic field of magnetic skyrmions due to their nontrivial  topology, the skyrmion Hall effect might be a roadblock for practical applications since any longitudinal motions of skyrmions in nanotrack is accompanied by a transverse motion.  A direct consequence of such an effect is easy destruction of skyrmions at the nanotrack edges during their fast motions along the nanotrack, despite their topological protection. Here we propose an entirely novel solution of completely inhibiting such skyrmion Hall effect without affecting its topological properties based on a antiferromagnetic-coupling bilayer system. We show that a pair of magnetic skyrmions can be nucleated in such a bilayer system through vertical current injection or converted from a current-driven domain-wall pair. Once nucleated, the skyrmion pair can be displaced through current-induced spin torque either from a vertical injected current or in-plane current. The skyrmion Hall effect is completely suppressed due to the cancellation of back-action forces acting on each individual skyrmion, resulting in a straight and fast motion of skyrmions along the current direction. This proposal will be of fundamental interests by introducing the bilayer degree of freedom into the system. Moreover, it provides an easy way to engineer the transport properties of the skyrmionic devices to achieve desired performance, making it highly promising for practical applications such as ultradense memory and information-processing devices based on skyrmions.
\end{abstract}
\maketitle


\address{{\normalsize Department of Applied Physics, University of Tokyo, Hongo 7-3-1, 113-8656, Japan}}

Since the first experimental observations of magnetic skyrmion lattices in
bulk non-centrosymmetric magnets\cite{Mol,Yu} and films\cite{Heinze}, there has been
tremendous interest in these topologically protected spin configurations
with a quantized topological number\cite{SkRev,Fert}. However, the creation and transmission
of isolated magnetic skyrmion in magnetic thin films are required for any
practical applications such as encoding information in individual skyrmion
to allow entirely novel devices and circuitry\cite{Sampaio,Fert,Tchoe,Yan,Marco,XichaoSR2015,XichaoSR2014}. Significant
efforts and progress have been made towards realizing such ultrathin film
based on perpendicularly magnetized magnetic layer$|$heavy metal structure to host a rich variety of chiral spin textures
including skyrmions\cite{vincent,klaui}. The strong spin orbit coupling at the interfaces
between the magnetic layer with perpendicular magnetic anisotropy (PMA) and
the underlying heavy metal layer provides a sizeable Dzyaloshinsky-Moriya
interaction (DMI) to stabilize skyrmions\cite{Fert,parkin,Bogdanov_1,Bogdanov_2,Bogdanov_3}.

To move a skyrmion in nanotrack for information-processing applications, a convenient and
efficient way is by means of spin current which can transfer the angular
momentum from itinerant conduction electrons to the magnetic moments of the
skyrmion\cite{IwasakiNL,Iwasaki}. However, one major roadblock to the manipulation and transmission
of skyrmions in nanotrack is the skyrmion Hall effect, \textit{i.e.},
skyrmions exhibit the Hall effect driven by spin currents due to the presence of
the Magnus force which in turn originates from its nontrivial topology\cite{SkRev}.
Thus a skyrmion will not move parallel with the direction of the current.
Instead it will gain a transverse velocity with a magnitude proportional to
the spin current density which displaces the skyrmion towards the edge of
the nanotrack. Therefore a skyrmion will be easily destroyed for a distance of
much less than 1$\mu $m for a nanotrack made of typical magnetic layer$|$heavy
metal system\cite{IwasakiNL,Iwasaki,IwasakiNC}.

\begin{figure*}[tbp]
\centerline{\includegraphics[width=0.96\textwidth]{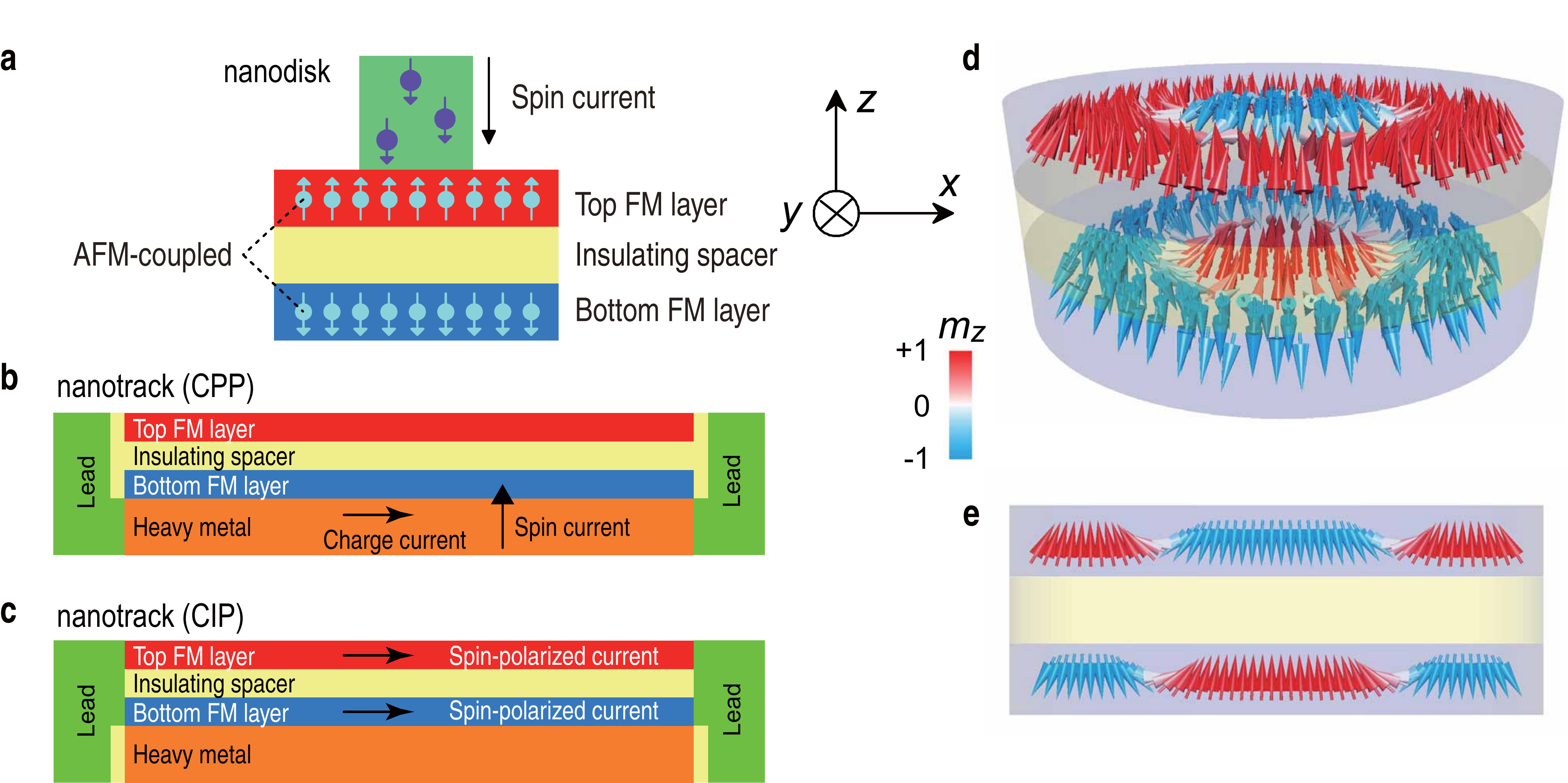}}
\caption{\textbf{Schematics of the bilayer nanodisk, nanotrack and the bilayer-skyrmion.}
\textbf{a}, The bilayer nanodisk for the creation of skyrmions, of which
the diameter is 100 nm. The spin current (polarized along $-z$) is injected
into the top layer in the central circle region with a diameter of 40 nm.
\textbf{b}, The bilayer nanotrack (500 nm $\times$ 50 nm $\times$ 3 nm)
for the motion of skyrmion driven by current perpendicular to the plane (CPP).
The charge current flows through the heavy metal along the $x$-direction, which
leads to the generation of spin current (polarized along $+y$) perpendicularly
injected to the bottom layer due to the spin Hall effect. The skyrmion in the
bottom layer is driven by the spin current, while the skyrmion in the top layer
moves remotely due to the interlayer exchange coupling.
\textbf{c}, The bilayer nanotrack (500 nm $\times$ 50 nm $\times$ 3 nm) for
the motion of skyrmion driven by in-plane current (CIP). The electrons flow
towards the right in both the top and bottom layers, $i.e.$, the charge currents
flow  along the -$x$-direction. The skyrmions in both the top and bottom
layers are driven by the spin current. In all models, the thickness of both the
top ferromagnetic (FM) layer, the bottom FM layer and the insulating spacer
are equal to 1 nm. The top layer and bottom layer is antiferromagnetically exchange-coupled, where the initial state of the top layer is almost spin-up (pointing
along +$z$) and that of bottom layer is almost spin-down (pointing along -$z$).
\textbf{d}, Illustration of the bilayer-skyrmion in a nanodisk, which is a set of
antiferromagnetically exchange-coupled skyrmions.
\textbf{e}, Sideview of the bilayer-skyrmion along the diameter of \textbf{d}. The
color scale represents the out-of-plane component of the magnetization, which is used throughout this paper.}
\label{Fig1}
\end{figure*}

Since both the skyrmion Hall effect and its topological protection arise
from the same mechanism, \textit{i.e}., its nontrivial topological number of
$\pm 1$, it seems to be impossible to inhibit the skyrmion Hall effect without
breaking its topological protection. In this work, we propose a novel
solution of two perpendicularly magnetized sublayers strongly coupled via
the antiferromagnetic (AFM) exchange interaction with a heavy metal
layer beneath the bottom magnetic layer\cite{ParkinNNano2015}
(Fig.\ref{Fig1}a, b, c).
When a skyrmion is created in one of the sublayers, another skyrmion is simultaneously created in the other sublayer
under certain conditions. We refer to such a pair of magnetic skyrmions as a magnetic bilayer-skyrmion
(Fig.\ref{Fig1}d, e).
Moreover, we show a bilayer-skyrmion can be displaced over arbitrarily long distances driven by spin currents without
touching the edges due to the absence of the skyrmion Hall effect, which is distinct from a skyrmion in the monolayer thin-film structure.
In addition to address the above-mentioned dilemma,
bilayer-skyrmions have many distinct characteristics from conventional skyrmions in monolayer thin film,
which allow for versatile and multifunctional ultradense and ultrafast information processing and logic applications.

These features are true for both the current-in-plane (CIP) and
current-perpendicular-to-plane (CPP) cases. The CIP implementation
has been extensively studied in the past ten
years for displacing domain walls or switching magnetization.
Recently, the spin Hall effect has been demonstrated
to be a more efficient means of manipulating magnetization.
The current for the CIP case is along
the nanowire axial direction. By contrast, the spin current is perpendicular to the heavy metal/ferromagnetic interface for the CPP case. In our
nomenclature for CPP, the spin current is perpendicular-to-film
plane whereas the charge current can be either in the film
plane (such as the spin Hall effect scenario) or perpendicular to
the film plane (such as the perpendicular MRAM where a perpendicularly
magnetized polarizer is incorporated). The CPP scheme has been
proven to be much more efficient to move the domain wall or skyrmion than the CIP method\cite{parkin,Sampaio,Iwasaki,IwasakiNL}.

In this work, we first describe the nucleation process of an isolated skyrmion within the top layer  by utilizing the CPP injection into a disk composed of two AFM-coupled magnetic sublayers. Owing to the interlayer AFM exchange coupling, another skyrmion will automatically emerge in the bottom layer. In so doing we explore various properties of a bilayer-skyrmion. Furthermore, we move a bilayer-skyrmion in the nanotracks either by the CPP or CIP. We also study the generation of a domain-wall (DW) pair and the conversion process from a DW pair to a skyrmion in the bilayer nanotracks.

\begin{figure*}[t]
\centerline{\includegraphics[width=0.9\textwidth]{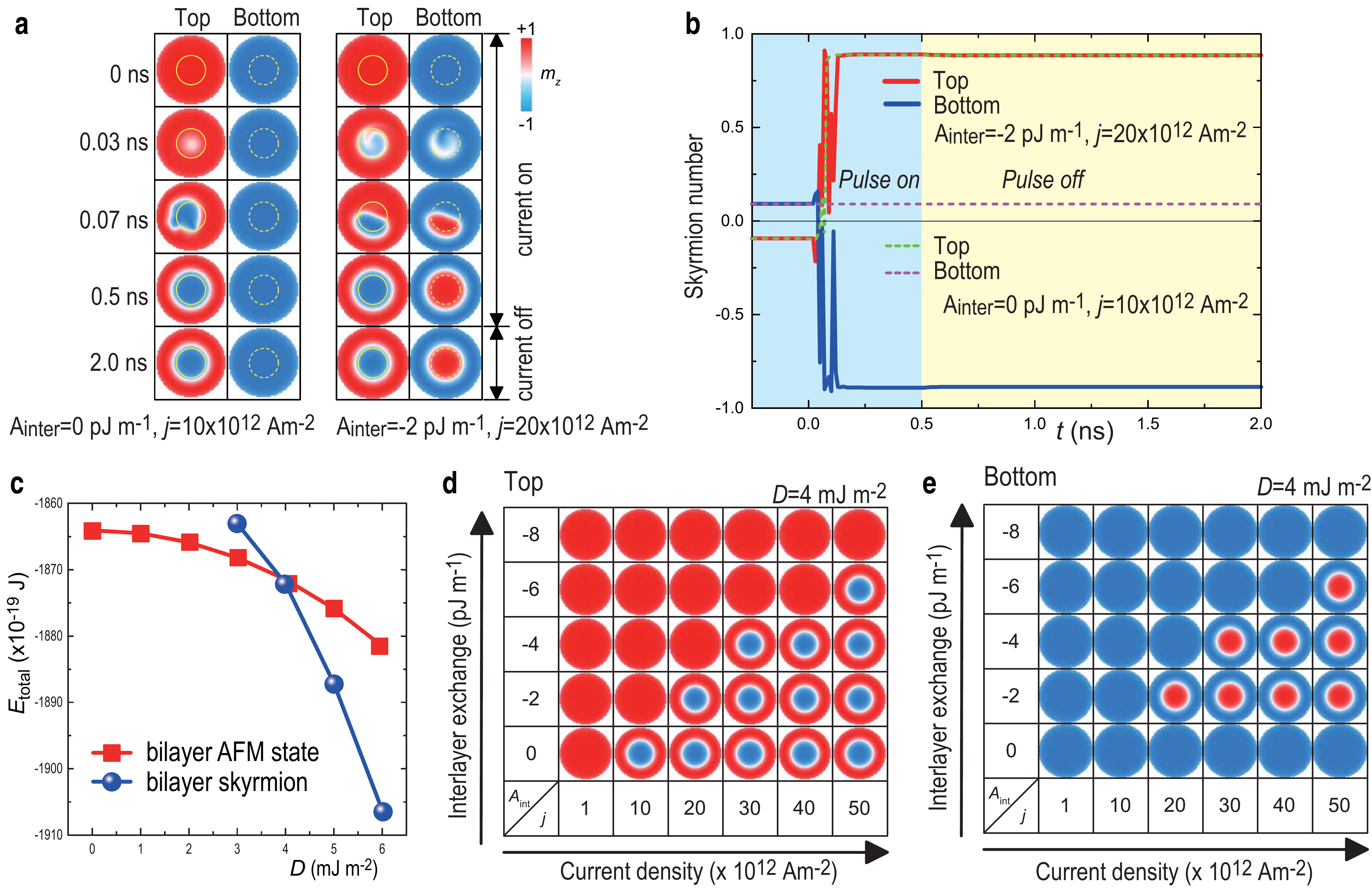}}
\caption{\textbf{Creation of skyrmions in the bilayer nanodisk and time evolution of the skyrmion number.}
\textbf{a}, Injection of skyrmions in the bilayer nanodisk ($D$ = 4 mJ m$^{-2}$) with/without
interlayer AFM exchange coupling. A 0.5-ns-long spin current ($P$ = 0.4) is injected
into the top layer (denoted by solid yellow circles), followed by a 1.5-ns-long relaxation.
The interlayer exchange constant $A_{\text{inter}}$ is set as 0 or -2 pJ m$^{-1}$, whereas
the corresponding interface exchange constant $\protect\sigma$ equals to 0 or -2 mJ m$^{-2}$ (\textcolor{black}{see Supplementary Movies 1-2}).
\textbf{b}, The time evolution of the skyrmion number of the top and bottom layers in
the nucleation process of skyrmions corresponding to \textbf{a}.
\textbf{c}, Total micromagnetic energy $E_{\text{total}}$ (including the intralayer exchange, interlayer exchange, dipolar,
 anisotropy and DMI energy) for a bilayer-skyrmion and the AFM-coupled ground state as a function of the DMI constant $D$.
Relaxed state of (\textbf{d}) the top layer and (\textbf{e}) the bottom layer after
the injection of a 0.5-ns-long spin current for various current density $j$ and interlayer
exchange constant $A_{\text{inter}}$.
}
\label{Fig2}
\end{figure*}

\section{Bilayer System coupled with AFM interaction}

\textbf{Hamiltonian.}
We investigate the bilayer system where the top and
bottom ferromagnetic (FM) layers are coupled antiferromagnetically by the
exchange interaction, as illustrated in Fig.\ref{Fig1}a. The Hamiltonian
for each layer reads
\begin{align}
H_{\tau }=&-A_{\text{intra}}\sum_{\langle i,j\rangle }\boldsymbol{m}_{i}^{\tau }
\cdot \boldsymbol{m}_{j}^{\tau }+\sum_{\langle i,j\rangle }
\boldsymbol{D}\cdot (\boldsymbol{m}_{i}^{\tau }\times \boldsymbol{m}_{j}^{\tau })\notag\\
&+K\sum_{i}[1-(m_{i}^{\tau ,z})^{2}]+H_{\text{DDI}},  \label{HamilAF}
\end{align}
where $\tau $ is the layer index ($\tau =$T, B), $\boldsymbol{m}_{i}^{\tau }$
represents the local magnetic moment orientation normalized as
$|\boldsymbol{m}_{i}^{\tau }|=1$ at the site $i$, and $\left\langle i,j\right\rangle $
runs over all the nearest neighbor sites in each layer. The first term
represents the FM exchange interaction with the FM exchange stiffness $A_{\text{intra}}$.
The second term represents the DMI with the DMI vector $\boldsymbol{D}$. The
third term represents the PMA with the
anisotropic constant $K$. $H_{\text{DDI}}$ represents the dipole-dipole interaction.
There exists an AFM coupling between the top and
bottom layers,
\begin{equation}
H_{\text{inter}}=-A_{\text{inter}}\sum_{i}\boldsymbol{m}_{i}^{\text{T}}\cdot
\boldsymbol{m}_{i}^{\text{B}}.
\end{equation}
The sign of $A_{\text{inter}}$ is negative reflecting that the interlayer
interaction is antiferromagnetic. We assume that the spins in the top layer
are pointing upward. Then the spins in the bottom layer are pointing downward due to
the interlayer AFM couplings.

\begin{figure*}[t]
\centerline{\includegraphics[width=0.96\textwidth]{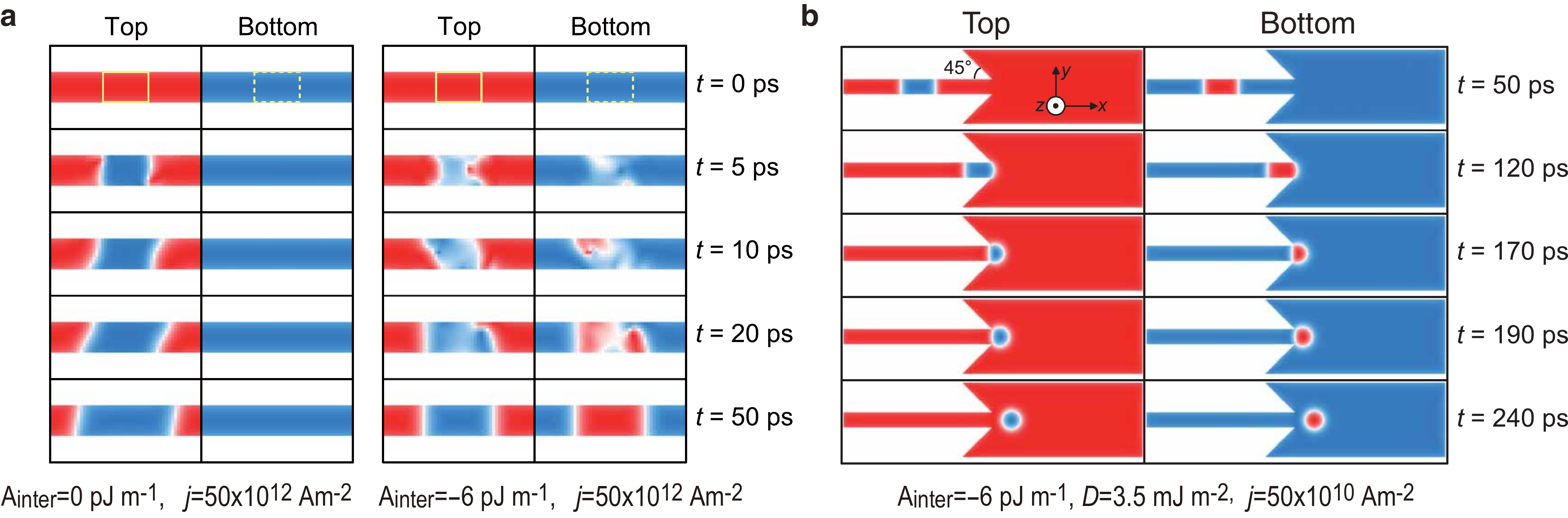}}
\caption{\textbf{Creation of a bilayer DW pair, its conversion into a bilayer-skyrmion,
and their motions driven by vertical current in a bilayer nanotrack.}
The length, wide width and narrow width of the nanotrack ($D$ = 3.5 mJ m$^{-2}$) equal 400 nm, 100 nm, and 20 nm, respectively.
\textbf{a}, A local vertical spin current ($j$ = 50 $\times$ 10$^{12}$ A m$^{-2}$, $P$ = 0.4, polarized along $-z$) is perpendicularly applied to the top layer of the narrow side (85 nm < $x$ < 115 nm) before $t$ = 50 ps. When the top and bottom layers are decoupled ($A_{\text{inter}}$ = 0 pJ m$^{-1}$), only one DW pair is generated in the top layer. However, when the top and bottom layers are coupled ($A_{\text{inter}}$ = -6 pJ m$^{-1}$), a bilayer DW pair is created at $t$ = 50 ps.
\textbf{b}, A global vertical spin current is perpendicularly applied to the bottom layer (towards +$z$, polarized along +$y$) when the bilayer DW pair is created at $t$ = 50 ps. The current density $j$ in wide part equals
5 $\times$ 10$^{11}$ A m$^{-2}$, which is proportional to that inside the narrow part with
respect to the ratio of narrow proportion (200 nm $\times$ 20 nm) to wide proportion
(200 nm $\times$ 100 nm). The 45-dgree notches are employed to reduce the required current
density for the DW-skyrmion conversion. When the global driving current is turned on at $t$ = 50 ps, the local DW injection current is turned off at the same time. \textcolor{black}{See Supplementary Movie 3.}}
\label{Fig3}
\end{figure*}

\textbf{Topological Number.}
The classical field $\boldsymbol{m}^{\tau }(\boldsymbol{x})$
is introduced for the spin texture in the FM system by considering the zero limit of the the lattice constant,
$a\rightarrow 0$. The ground-state spin textures are
$\boldsymbol{m}^{\text{T}}=(0,0,1)$ and $\boldsymbol{m}^{\text{B}}=(0,0,-1)$. A magnetic skyrmion is
a spin texture which has a quantized topological number. Spins swirl
continuously around the core and approach the ground-state value
asymptotically. The skyrmion is characterized by the topological number $Q_{\tau }$ in each layer,
\begin{equation}
Q_{\tau }=-{\frac{1}{4\pi }}\int d^{2}x\boldsymbol{m}^{\tau }(\boldsymbol{x})
\cdot \left( \partial _{x}\boldsymbol{m}^{\tau }(\boldsymbol{x})\times
\partial _{y}\boldsymbol{m}^{\tau }(\boldsymbol{x})\right) .
\label{PontrNumbe}
\end{equation}
We obtain $Q_{\tau }=\pm 1$ for a skyrmion in a sufficiently large area.
We also call $Q_{\tau }$ the skyrmion number.
Even if the skyrmion spin texture is deformed, its skyrmion number does not change,
as far as the boundary condition is not modified. It can be
neither destroyed nor separated into pieces, \textit{i.e.}, a skyrmion is
topologically protected.

The spins in the top and bottom layers are tightly bounded due to the AFM
coupling. Accordingly, if one skyrmion is created in the top layer, a second skyrmion
is also created in the bottom layer (Fig.\ref{Fig2}a) simultaneously.
The topological number of the bottom layer is opposite to that of the top layer since all the spins are
inverted, $Q_{\text{B}}=-Q_{\text{T}}$.
See Fig.\ref{Fig2}b how these topological numbers evolve after the creation of a skyrmion in the top layer.

\textbf{Excitation Energy.}
We may determine numerically the spin profile of
each skyrmion and estimate the excitation energy based on the Hamiltonian
$H_{\text{total}}=H_{\text{T}}+H_{\text{B}}+H_{\text{inter}}$.
We compared the energy of one bilayer-skyrmion state and the energy of the homogeneous state in Fig.\ref{Fig2}c as
a function of the DMI. Similar result is obtained for any value of the interlayer AFM coupling strength.
This is because the spin directions between the top and bottom layers are always opposite.

The energy of the bilayer-skyrmion becomes lower than that of the uniform ground state for $D\gtrsim 4$ mJ m$^{-2}$,
which is the threshold value.
Consequently, the bilayer-skyrmion excitation is energetically favorable
when the DMI strength is larger than the threshold value.
However, spontaneous generation of skyrmions does not occur due to the topological protection.
Nevertheless, the topological protection can be violated in condensed matter
physics because of the existence of the lattice structure and the boundary
of the sample. We may create skyrmions by leveraging these properties.

\textbf{Skyrmion Hall effect.}
It is well understood\cite{Stone} that
the center-of-mass motion of a skyrmion is determined by the Lagrangian,
\begin{equation}
L=L_B-U,
\end{equation}
where $L_B$ is the Berry phase term of the spin texture,
\begin{equation}
L_B=\frac{G}{2}(\dot{X}Y-X\dot{Y})=\frac{1}{2}\mathbf{G}\cdot (\dot{\mathbf{R}}\times\mathbf{R}),
\end{equation}
and $U$ is the potential.
Here,  $\mathbf{G}=(0,0,G)$ is the gyromagnetic coupling constant representing the Magnus force with $G=4\pi Q$,
and $\mathbf{R}=(X,Y)$ is the center-of-mass coordinate of the skyrmion.
The Euler-Lagrange equation yields
\begin{equation}
\dot{\mathbf{R}}=\frac{1}{G}\mathbf{e}_z\times \mathbf{F},
\end{equation}
where $\mathbf{e}_z=(0,0,1)$, and $\mathbf{F}=-\nabla U$ is the force acting on the skyrmion.
Consequently, a moving skyrmion feels the Magnus force and bends toward the direction perpendicular to the force.
It is called the skyrmion Hall effect.

The direction of the Magnus force is opposite when the sign of the skyrmion number $Q$ is opposite.
For instance, if a skyrmion is driven by the current along the $+ x$ direction it will be bended toward the $+ y$ ($-y$) direction in the top (bottom) layer.

\section{Landau-Lifshitz-Gilbert-Slonczewski equation for CPP}

We may apply a CPP spin-polarized current injection from magnetic tunnel junction (MTJ) or the spin Hall effect in heavy metal layer\cite{Khv,Sampaio,Fino}.
We numerically solve the Landau-Lifshitz-Gilbert-Slonczewski (LLGS) equation, which governs the
dynamics of the magnetization $\boldsymbol{m}_{i}$ at the lattice site $i$. By suppressing the layer index, it reads
\begin{align}
\frac{d\boldsymbol{m}_{i}}{dt} =&-|\gamma |\boldsymbol{m}_{i}\times
\boldsymbol{H}_{i}^{\text{eff}}+\alpha \boldsymbol{m}_{i}\times
\frac{d\boldsymbol{m}_{i}}{dt}  \notag \\
& +\left\vert \gamma \right\vert u(\boldsymbol{m}_{i}\times \boldsymbol{p}\times \boldsymbol{m}_{i})
-\left\vert \gamma \right\vert u^{\prime }(\boldsymbol{m}_{i}\times \boldsymbol{p}),  \label{LLGS}
\end{align}where $\boldsymbol{H}_{i}^{\text{eff}}\mathbf{=-}
\partial H_{\text{total}}/\partial \mathbf{m}_{i}$ is the effective magnetic field induced by the
Hamiltonian $H_{\text{total}}=H_{\text{T}}+H_{\text{B}}+H_{\text{inter}}$, $\gamma $ is the
Gilbert gyromagnetic ratio, $\alpha $ is the Gilbert-damping coefficient
originating from spin relaxation, $u$ is the Slonczewski STT coefficient,
$u^{\prime }$ is the out-of-plane STT coefficient, and $\boldsymbol{p}$
represents the electron polarization direction. Here, $u=|\frac{\hbar }{\mu
_{0}e}|\frac{j|\boldsymbol{p}|}{2dM_{s}}$ with $\mu _{0}$ the vacuum
magnetic permittivity, $d$ the film thickness, $M_{s}$ the saturation
magnetization, and $j$ the current density. We take $-z$ direction for
creating the skyrmion, while $+y$ direction for moving the skyrmion.
The STT is induced either by injection from a magnetic tunnel junction polarizer or
by the spin Hall effect\cite{Sampaio,Fino,Khv}.

\begin{figure*}[t]
\centerline{\includegraphics[width=0.94\textwidth]{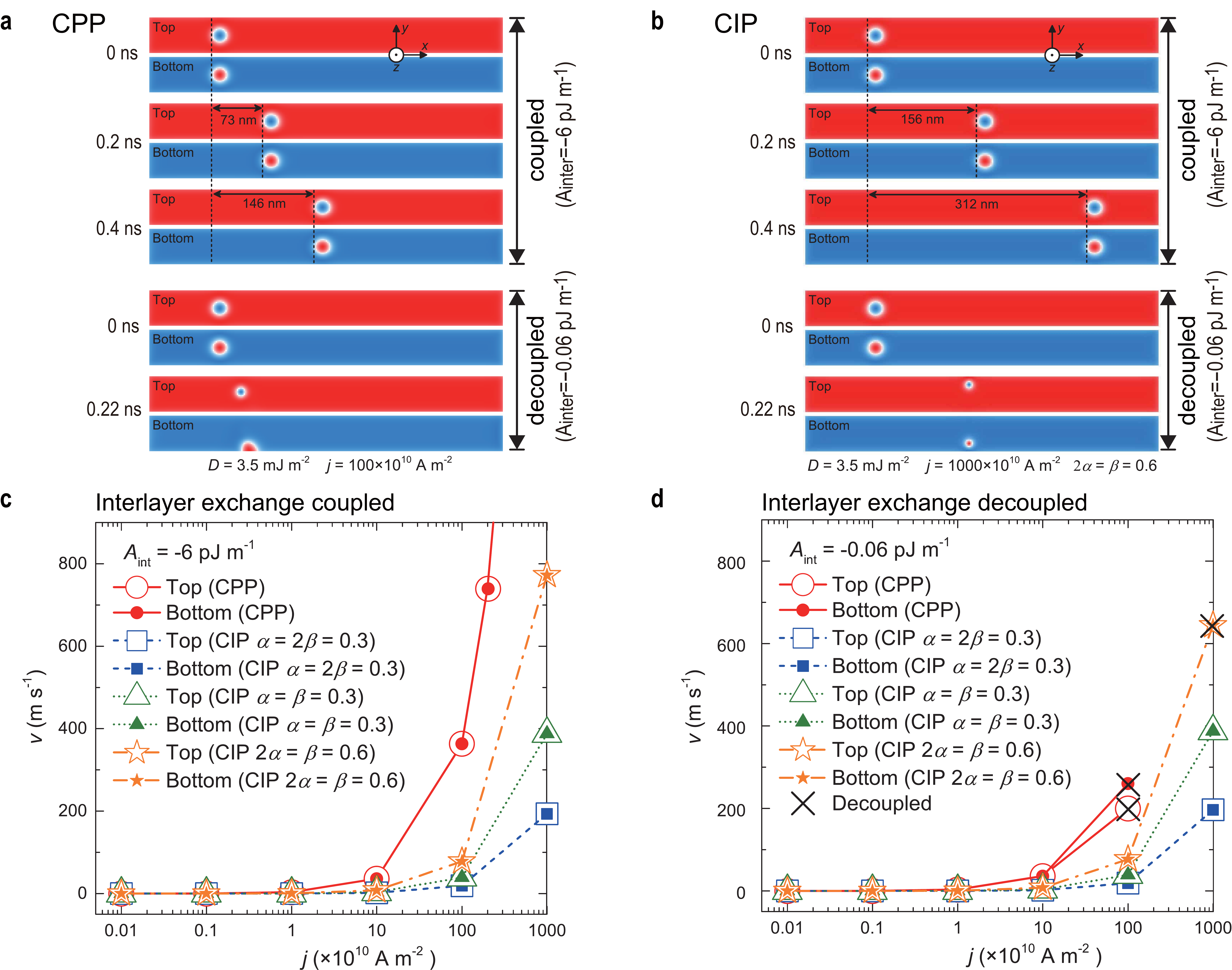}}
\caption{\textbf{Motion of skyrmions in the top and bottom layers of a bilayer nanotrack.}
Top-views of motion of skyrmions at selected interlayer exchange coupling and times driven by
\textbf{(a)} the CPP injection and \textbf{(b)} the CIP injection (\textcolor{black}{see Supplementary Movies 4-7}). The
parameters of the bilayer nanotrack are 500 nm $\times$ 50 nm $\times$ 3 nm, $D$ = 3.5 mJ m$^{-2}$. The skyrmions are
initially created by MTJ skyrmion injector placed on the top layer at $x$ = 100 nm. For the CPP case,
the spin current ($P$ = 0.4) in the bottom layer is applied along +$z$ but polarized along +$y$.
The skyrmion in the bottom layer moves towards right driven by the spin current, while the skyrmion
in the top layer remotely moves due to the interlayer coupling. For the CIP case, the skyrmions in
both the top and bottom layers are driven by in-plane spin currents ($P$ = 0.4). The velocities of
skyrmions in the top and bottom layers as functions of current density $j$ with \textbf{(c)} large
interlayer exchange $A_{\text{inter}}$ = -6 pJ m$^{-1}$ and \textbf{(d)} small interlayer exchange
$A_{\text{inter}}$ = -0.06 pJ m$^{-1}$. The cross symbol denotes the decoupling and destruction of
skyrmions in the top and bottom layers due to large current density and small interlayer exchange
coupling, where the velocities are calculated before the destruction of skyrmion.}
\label{Fig4}
\end{figure*}

\section{LLG equation for CIP}

Alternatively we may apply a CIP injection\cite{Khv,Sampaio,Fino} to move skyrmions.
We numerically solve the LLG equation,
\begin{align}
\frac{d\boldsymbol{m}_{i}}{dt} =&-|\gamma |\boldsymbol{m}_{i}\times
\boldsymbol{H}_{i}^{\text{eff}}+\alpha \boldsymbol{m}_{i}\times \frac{d\boldsymbol{m}_{i}}{dt}  \notag \\
&+\frac{\boldsymbol{p}|a^{3}}{2eM_{s}}(\boldsymbol{j(r)}\cdot \nabla )\boldsymbol{m}_{i} \notag \\
&-\frac{\boldsymbol{p}|a^{3}\beta }{2eM_{s}}[\boldsymbol{m}_{i}\times (\boldsymbol{j(r)}\cdot \nabla )\boldsymbol{m}_{i}],
\end{align}
where $\beta$ is the strength of the non-adiabatic torque and $a$ is the lattice constant.

\textbf{Creation of a bilayer-skyrmion by vertical spin current.}
We employ a
CPP injection with a circular geometry in a nanodisk.
The spin-polarized current (polarized along $-z)$ is injected into the top layer in the central circle
region, as illustrated in Fig.\ref{Fig1}a.

We demonstrate how the spin textures develop in Fig.\ref{Fig2}a.
The spins start to flip in both layers following the spin current injection only in the top layer.
When there is no interlayer AFM coupling, a skyrmion is formed only in the top
layer (\textcolor{black}{see Supplementary Movie 1}). By contrast, a skyrmion is formed
also in the bottom layer upon the current injection in the presence of the interlayer AFM coupling (\textcolor{black}{see Supplementary Movie 2}).

We show the evolution of the skyrmion number in Fig.\ref{Fig2}b.
It oscillates at the initial stage for $t<0.2$ ns, and rapidly increases to $1$.
The skyrmion remains stable even when the current is switched off, demonstrating that it is topologically protected.
During this process, the spins in the top and bottom layers are always anti-parallel.
A comment is in order.
The saturated skyrmion number is not exactly $Q=1$ but $Q=0.89$.
This is due to the fact that there is a background skyrmion number which originates from the tilting the edge spins. It is $Q=-0.09$.
Accordingly, the calibrated skyrmion number is $Q=0.98$, which is almost unity.

We present a nucleation phase diagram of a bilayer-skyrmion pair as a function of the current density
and the interlayer AFM coupling in Fig.\ref{Fig2}d and Fig.\ref{Fig2}e.
When the magnitude of the injected current is strong enough, the bilayer-skyrmion is created.
This is due to the fact that spin flip costs a certain energy.
On the other hand, if the interlayer AFM coupling is too strong,
the bilayer-skyrmion is suppressed due to the fact that the nucleation field and the
coercivity increases with the interlayer AFM exchange, leading to a larger
current density for nucleation.

\textbf{Creation of a bilayer-skyrmion from a bilayer DW pair.}
A magnetic skyrmion can be created from a DW pair by using a junction geometry\cite{NC}.
In this scenario we first make a DW pair into a nanotrack of the top layer through the local CPP injection with $-z$ direction.
We show how the spins start to flip in the top layer and subsequently in the bottom layer driven by the AFM exchange force in Fig.\ref{Fig3}a.
Then, the bilayer DW pair is shifted by applying CPP current,
as shown in the process from $t=50$ ps to $t=120$ ps in Fig.\ref{Fig3}b.
Here we consider the vertical injection of a spin current towards $+z$ and polarized along $+y$ in the bottom layer.
The CPP injection moves the bilayer DW in the rightward direction. When the
bilayer DW arrives at the junction interface ($t=170$ ps), both the end spins
of the DW are pinned at the junction, whereas the central part of the DW
continues to move due to STT in the wide part of the nanotrack. Therefore,
the structure is deformed into a curved shape and a bilayer-skyrmion
texture forms at $t=190$ ps (\textcolor{black}{see Supplementary Movie 3}).

\textbf{Current-driven motion of a bilayer-skyrmion in a nanotrack.}
The magnetic bilayer-skyrmion can be displaced by the vertical spin-polarized current as in the case of the magnetic skyrmion.
We may employ the CPP injection or the CIP injection to drive a bilayer-skyrmion.
In general, a moving skyrmion is easily destroyed by the sample edges due to the skyrmion Hall effect.
Therefore, the maximum velocity of skyrmion in FM nanotrack is
typically much less than 10$^{3}$ m/s, limited by the edge confining force\cite{IwasakiNL} of $\sim D^{2}/J$.

The skyrmion in the top layer follows the motion of the skyrmion in the bottom layer even when the current is not injected into the top layer.
This is because that two skyrmions are bounded by the interlayer AFM coupling.
There is no skyrmion Hall effect for a magnetic bilayer-skyrmion.
This can be explained as follows.
If there is no interlayer AFM coupling, a skyrmion in the top layer
moves left-handed and the skyrmion in the bottom layer moves right-handed. However,
when the AFM coupling is strong enough, two skyrmions are tightly bounded and the Magnus forces acting on the skyrmions between the top and bottom layers are exactly cancelled.
Accordingly the bilayer-skyrmion will move straightly.
This mechanism works  both for the CPP and CIP cases (\textcolor{black}{see Supplementary Movies 4-5}).

We show the relation between the magnitude of the injected current and the
velocity in Fig.\ref{Fig4}. The velocity is proportional to the injected current density. For strong enough current, the bilayer-skyrmion is
destroyed and split into two independent skyrmions. This is because that
the skyrmion Hall effect increases as the current increases, which acts as
the repulsive force between two skyrmions (\textcolor{black}{see Supplementary Movies 6-7}).

With strong interlayer AFM exchange coupling, the coupled
skyrmions move along the central line of the nanotrack at a high speed of
a few hundred meters per second, without any transverse motion. However, with
small interlayer AFM exchange coupling, the skyrmions in the top
and bottom layers will be decoupled due to the fast motion of the
skyrmion in the bottom layer driven by large current. Once the skyrmions in the top
and bottom layers are decoupled, the skyrmion Hall effect becomes active,
leading to the destruction of skyrmions in the top and/or bottom layer by
touching edge. At the same time, in the CPP case, the skyrmion in the top layer will
stop motion. It can be seen that the coupled skyrmions driven by CIP current
doesn't require a fine tuning of damping and non-adiabatic torque
coefficients.

It is also worth noting that the interlayer AFM coupling does not produce a mass of the bilayer-skyrmion. When the driving current is suddenly turned off, the bilayer-skyrmion stops high-speed motion immediately (see Supplementary Movie 8). On the other hand, we also investigated the case where the DMI constant $D$ is different between the top and bottom layers. It is found that the results of current-driven motion of the bilayer-skyrmion do not change much since the DMI only changes the radius of the skyrmion (see Supplementary Movies 9-10). The massless property and its robustness make the bilayer-skyrmion an ideal candidate for practical applications.

\section{Perspectives}
We have presented a novel solution of inhibiting the Hall effect of skyrmions without affecting their topological properties,
by exploring a new device made of antiferromagnetically exchange-coupled bilayer nanodisks and nanotracks.
Compared with the mostly investigated skyrmion in the FM monolayer
system, the bilayer-skyrmion exhibits entirely distinct characteristics
with regard to the current-transport behaviour and
robustness. First, it can move strictly along the direction of
the spin current flow, which makes it more appealing for motions
in nanowires for ultradense memory applications. This
is in high contrast with the case of monolayer skyrmion, where the skyrmion information carrier can be easily
destroyed by the edges of the nanotracks. Second, it will be
immune to magnetic field perturbations which might be generated
externally or internally within the device circuitry since
the net magnetic moment is zero. Third, by introducing
the bilayer degree of freedom,
the transport properties of the device can be engineered to achieve desired performance. For
example, the skyrmion Hall effect can be intentionally suppressed
or enhanced by tuning the magnetic properties of individual layers.
This newly proposed solution of transporting
skyrmion information carrier for arbitrarily long distances
at much enhanced velocity may be very appealing for versatile
applications such as ultradense memory and information
processing. Similar ideas can be extended to multilayer or superlattice
where the skyrmions are strongly coupled to realize
a better manipulation of skyrmions in nanotrack or extended
thin films.

\section{Methods}

\textbf{Modeling and simulation.} The micromagnetic
simulations are performed using the Object Oriented MicroMagnetic Framework
(OOMMF) including the Dzyaloshinskii-Moriya interaction (DMI) extended module
\cite{Fert, Sampaio, Dona, Rohart}. The time-dependent magnetization dynamics
is governed by the Landau-Lifshitz-Gilbert (LLG) equation including spin
torque\cite{Brown, Gilbert, LL, Thia, Thia2}. The average energy density $E$ is a function of $\textbf{M}$, which contains the
intralayer exchange, the interlayer exchange, the anisotropy, the applied
field (Zeeman), the demagnetization and the DMI energy terms. For
micromagnetic simulations, the intrinsic magnetic parameters are adopted
from Refs.\cite{Fert, Sampaio}: Gilbert damping coefficient $\alpha
= 0.3$ and the value for Gilbert gyromagnetic ratio is -2.211$\times$ 10$^{5}$ m A$^{-1}$ s$^{-1}$.
Saturation magnetization $M_{S}$ = 580 kA m$^{-1}$, intralayer
exchange stiffness $A$ = 15 pJ m$^{-1}$, DMI constant $D$ = 0 $\sim$ 6 mJ m$^{-2}$ and
perpendicular magnetic anisotropy (PMA) $K$ = 0.8 MJ m$^{-3}$ unless otherwise
specified. The interlayer exchange coefficient $A_\text{inter}$ is set from 0 to -10 pJ
m$^{-1}$, whereas the corresponding interface exchange coefficient $\sigma$
equals from 0 to -10 mJ m$^{-2}$ ($\protect\sigma$ = $A_{\text{inter}}$ / 1 nm),
where "-" denotes that the interface is
antiferromagnetically coupled. The field-like out-of-plane STT coefficient $u^{\prime }$
is set to zero. All samples are discretized into cells of 2 nm $\times$ 2 nm $\times$ 1 nm
in the simulation, which is sufficiently smaller than the typical exchange length
($\sim$ 4.3 nm) and the skyrmion size to ensure the numerical accuracy.
For all simulation of current-driven skyrmions reported throughout this paper, the skyrmions
are firstly created at the designed spot of the nanotrack ($x$ = 100 nm) by a local spin current
perpendicular to the plane of the top layer. Then the system is relaxed to an
energy minimum state without applying any current. Next, we start the timer
and the spin current ($P$ = 0.4) is injected into the nanotrack with the
geometry of current-in-plane (CIP) or current-out-of-plane (CPP) as shown in Fig.\ref{Fig1}.
In the configuration of CIP, the electrons flow toward the right in both the top and
bottom layers, $i.e.$, the currents flow toward the left, while in the
configuration of CPP, the electrons flow toward the top only in the bottom
layer.

\section{Acknowledgements}
Y.Z. thanks the support by the Seed Funding Program for Basic Research and
Seed Funding Program for Applied Research from the HKU, ITF Tier 3 funding
(ITS/171/13), the RGC-GRF under Grant HKU 17210014, and University Grants
Committee of Hong Kong (Contract No. AoE/P-04/08). M.E. thanks the support
by the Grants-in-Aid for Scientific Research from the Ministry of Education,
Science, Sports and Culture, No. 25400317. M.E. is very much grateful to N.
Nagaosa for many helpful discussions on the subject.

\section{Author contributions}
M.E. conceived the idea and designed the project.
M.E. and Y.Z. coordinated the project.
X.Z. performed the numerical simulations supervised by Y.Z.
All authors discussed the results and wrote the manuscript.

\section{Additional information}
Supplementary information is available.

\section{Competing financial interests}
The authors declare no competing financial interests.

\end{document}